\documentclass[11pt]{article}
\usepackage{graphicx}
\usepackage{subfigure}
\begin{document}
\oddsidemargin .03in
\evensidemargin 0 true pt
\topmargin -.6in

%Abbreviations %
%***************%

\def\ra{{\rightarrow}}
\def\a{{\alpha}}
\def\b{{\beta}}
\def\l{{\lambda}}
\def\eps{{\epsilon}}
\def\T{{\Theta}}
\def\t{{\theta}}
\def\co{{\cal O}}
\def\car{{\cal R}}
\def\caf{{\cal F}}
\def\cs{{\Theta_S}}
\def\pr{{\partial}}
\def\tri{{\triangle}}
\def\na{{\nabla }}
\def\S{{\Sigma}}
\def\s{{\sigma}}
\def\sp{\vspace{.15in}}
\def\hs{\hspace{.25in}}

\newcommand{\be}{\begin{equation}} \newcommand{\ee}{\end{equation}}
\newcommand{\bea}{\begin{eqnarray}}\newcommand{\eea}
{\end{eqnarray}}

%********************************************************************%

\begin{titlepage}
\topmargin= -.4in
\textheight 9in

%\begin{flushright}
%{hep-th/0501}
%\end{flushright}
\begin{center}
\baselineskip= 18 truept

\vspace{.3in}
\centerline{\Large\bf Quantum Kerr tunneling vacua on a ${\mathbf{(D{\bar D})_4}}$-brane:}
\centerline{\Large\bf An emergent Kerr black hole in five dimensions}
\vspace{.6in}
\noindent
{{\bf Sunita Singh}, {\bf K.Priyabrat Pandey}, {\bf Abhishek K. Singh} {\bf and} {{\bf Supriya Kar}\footnote{skkar@physics.du.ac.in }}}

\vspace{.2in}
\noindent
{{\Large Department of Physics \& Astrophysics}\\
{\Large University of Delhi, New Delhi 110 007, India}}

\vspace{.2in}

{\today}
\thispagestyle{empty}

\vspace{.6in}
\begin{abstract}

\baselineskip=14 truept

\vspace{.12in}
We revisit a non-perturbative space-time curvature theory, underlying a two form $U(1)$ gauge dynamics, on a $D_4$-brane. In particular, two different gauge choices for a two form are explored underlying the dynamics of a geometric torsion in a second order formalism. We obtain two non-extremal quantum Kerr geometries in five dimensions on a pair of $(D{\bar D})_4$-brane in a type IIA superstring theory. The quantum vacua are described by a vanishing torsion in a gauge choice, underlying a geometric realization, on a non-BPS brane. It is argued that the quantum Kerr vacua undergo tunneling and lead to a five dimensional Kerr black hole in Einstein vacuum. A low energy limit in the quantum Kerr vacua further re-assures the emergent Kerr black hole.

\vspace{1in}
%\keywords {Near horizon D-brane, de Sitter tunnel tunnelling, Emergent gravity, AdS brane, String theory, Torsion geometry.}
%\arxivnumber{1303.4344}
\noindent
%{\sc Published in:} {\bf JHEP 05 (2013) 033}

\noindent
%{\sc ArXiv:} {\tt 13034344} {[hep-th]}
\end{abstract}
\end{center}

\vspace{.2in}

%\noindent PACS: 11.25.-w; 11.15.-q

%\noindent Keywords: String theory, D-branes, Noncommutative space-time and Gravitation with Torsion

%\thispagestyle{empty}%
\end{titlepage}

\baselineskip= 18 truept

%%%%%%%%%%%%%%%%%%%%%%
\section{Introduction}
%%%%%%%%%%%%%%%%%%%%%% 
The holographic idea \cite{susskind}, underlying bulk/boundary correspondences, has been established as a powerful tool to explore 
certain aspects of quantum gravity through a gauge theory or vice-versa. It is believed that the bulk physics underlying quantum gravity may emerge from
a boundary gauge theory. An emergent gravity has also been conjectured to arise due to the statistical behavior of microscopic degrees encoded on a holographic screen \cite{verlinde}. In fact, an emergent gravity has been a recent topic of great interest in the folklore of theoretical physics \cite{carlip,padmanabhan}. In the context, an emerging notion of quantum gravity in a gauge theory on a $D$-brane may turn out to be a potential candidate in string theory. 

\sp
\noindent
In the recent past, there have been attempts to construct extremal geometries on a BPS $D$-brane in various dimensions \cite{gibbons}-\cite{kahle-minasian}. They correspond to the near horizon black holes, primarily described in a ten dimensional type IIA or type IIB superstring theory.  For instance, a vacuum geometry in an open string bulk approaches its boundary at the black hole horizon and may be identified with a near horizon geometry. On the other hand, a BPS D-brane is described by a non-linear gauge theory and the non-linear electric charge has been pioneered to govern an extremal black hole. The emerging geometries on a $D$-brane, alternately underlie a non-commutative space-time on its world-volume. In a different context, a non-commutative space-time has also been explored to address an emergent gravity in a $U(1)$ gauge theory \cite{yang}-\cite{cai}.

\sp
\noindent
Importantly, a non-linear $U(1)$ on a single $D$-brane may be viewed through a modification of the linear $U(1)$ symmetry in presence of a zero mode of NS-NS two form. In other words, a zero mode in the string bulk couples to an electromagnetic field on a $D$-brane to form a non-linear $U(1)$ gauge invariant combination \cite{seiberg-witten}. However, the non-zero mode remains in the string bulk and hence does not play any significant role on a BPS D-brane. In the context, a non-zero mode has also been conjectured to describe an arbitrary non-commutative parameter on a $D$-brane and hence is believed to modify the curvature in the theory. Presumably a non-zero mode, of the NS-NS two form, has not been explored to its full strength in an open string world-sheet. Hence a non-zero mode remains a subtle issue from the perspective of an open string world-sheet. Nevertheless, a closed string world-sheet conformal symmetry in presence of the non-zero modes of a two form lead to the vanishing of beta function equations and re-assures a superstring effective action in ten dimensions. Thus, the non-zero modes do propagate in an effective space-time theory of a closed string. Generically, their dynamics do not seem to influence the non-linear gauge theory on an arbitrary $D$-brane. However on a $D_4$-brane, the non-linear $U(1)$ gauge dynamics may seen to be described by the non-zero modes of a two form. They lead to a generalized curvature formulated on an effective $D_4$-brane, in a second order formalism \cite{spsk}-\cite{kpss1}. 

\sp
\noindent
On the other hand, a two form gauge theory is Poincare dual to an one form dynamics a $D_4$-brane. The local degrees, of a two form, have been exploited on a $D_4$-brane to construct a geometric torsion non-perturbatively.  In fact, a geometric torsion has been constructed through a modification of the covariant derivative in presence of two form connections in a non-linear gauge theory. The iterative corrections in two form gauge connection lead to an exact description in a perturbative gauge theory. This, in turn, may be viewed as a non-perturbative geometric construction in a second order formalism. A priori, a geometric torsion breaks the $U(1)$ gauge invariance in the gauge theory. However, the gauge invariance is restored in a generalized curvature theory with the help of an emergent metric fluctuation. The local degrees of two form sources the fluctuations and they incorporate the notion of a dynamical space-time curvature on a $D_4$-brane. The emerging space-time may also be viewed through a two form gauge connection which takes an analogues form to the Christoeffel metric connection in Riemannian geometry. In a gauge choice for a two form, leading to a non-propgating torsion, the emerging fourth order curvature tensor reduces to a Riemannian curvature \cite{spsk-RN}. It may imply a generalization of Riemannian tensor to a special Cartan tensor, which asserts a generalized notion of emergent space-time described by a geometric torsion.

\sp
\noindent
One of the important aspects of the formalism lies in the non-extremal nature of the emergent geometries. It is important to note that the geometries are sourced by a non-trivial energy-momentum tensor underlying a two form gauge theory. The emergent geometries have been analyzed to address some of the non-extremal vacua in Einstein gravity in five dimensions. An extra stringy dimension, transverse to a BPS brane, presumably connects an anti BPS brane and together they lead to a non-BPS brane \cite{sen1}-\cite{sen2} in a global scenario. Since the formalism evolves with an intrinsic torsion, a $(D{\bar D})$-brane is inevitable to nullify the torison while establishing a correspondence with a known vacuum in Einstein gravity. However, the black hole mass generated by the torsion remains non-trivial in the global scenario described by a pair of $(D{\bar D})$-brane. In the recent past, the emergent geometries have been argued to describe a large density of tunneling vacua including de Sitter and anti de Sitter black holes \cite{spsk}. The large density of emergent vacua may be viewed via gauge transformations of two form. Presumably, they correspond to the landscape vacua in string theory \cite{quevedo,friedmann-stanley,louis}. 

\sp
\noindent
In the paper, we construct two emergent Kerr geometries leading to a quantum black hole in five dimensions on a non-BPS brane. We show that 
the quantum Kerr vacua in a low energy limit describes a five dimensional Kerr black hole \cite{myers-perry}-\cite{emparon-etal} in Einstein gravity underlying a closed string theory. A vanishing torsion, in both the gauge choices for a two form, ensures Riemannian geometry and a vacuum $T_{\mu\nu}=0$ solution in a generalized curvature formulation. It is argued that the background fluctuations in $B_2$ on a $D$-brane may have their origin in a dynamical two form described by a ten dimensional effective closed string action. In particular, one of the two quantum vacua in a low energy limit may a priori be viewed as a stringy Kerr vacuum. Nevertheless, with a subtlety, the low energy stringy Kerr vacuum may seen to characterize a Kerr black hole in Einstein vacuum. The limit truncates both the quantum Kerr geometries in the non-perturbative framework. On the other hand, the extra stringy dimensions in the framework may be scaled down with the $\t$-slicing geometries. They help to identify a stringy Kerr vacuum with a five dimensional Kerr black hole for a fixed polar angle. Analysis presumably suggests a tunneling between various quantum vacua on a non BPS brane. The quanta of radiations from the tunneling processes would like to lower the energy of the nucleated vacuum which in turn may lead to Einstein vacuum. In the context, the wall crossings analysis \cite{denef}-\cite{moore-etal} may provide hint to understand the transitions between some of the non-perturbative braneworld black holes. In other words, an exact geometry in perturbation theory validates a non-perturbative construction realized through a geometric torsion on a non-BPS brane. 

\sp
\noindent
We plan to begin with a moderate introduction to a non-perturbative geometric construction leading to a generalized curvature theory in five dimensions on a non BPS brane in section 2. Five dimensional quantum Kerr geometries in a generalized curvature theory, sourced by two different gauge choices for a two form are, respectively, worked out in section 3 and in section 4. We conclude by summarizing the results obtained in quantum gravity and an outlook beneath future research within the scope of the framework in section 5.

%%%%%%%%%%%%%%%%%%%%%%%%%%%%%%%%%%%%%%%%%%%%%%%%%%%%%%%%%%%%%
\section{Preliminaries: Setup on a $D_4$-brane}
%%%%%%%%%%%%%%%%%%%%%%%%%%%%%%%%%%%%%%%%%%%%%%%%%%%%%%%%%%%%%

%%%%%%%%%%%%%%%%%%%%%%%%%%%%%%%%%%%%%%
\subsection{A perturbative setup: Gauge theoretic curvature}
%%%%%%%%%%%%%%%%%%%%%%%%%%%%%%%%%%%%%%
A BPS brane carries an appropriate RR-charge and is established as a non-perturbative dynamical object in a ten dimensional type IIA superstring theory. In particular, a $D_4$-brane is governed by a supersymmetric gauge theory on its five dimensional world-volume. However, we restrict to the bosonic sector and begin with the $U(1)$ gauge dynamics, in presence of a constant background metric $g_{\mu\nu}$, on a $D_4$-brane. A linear one form dynamics is given by
\be
S_{\rm A}= -{1\over{4C_1^2}}\int d^5x\ {\sqrt{-g}}\ F^2\ ,\label{gauge-1}
\ee 
where $C_1^2=(4\pi^2g_s){\sqrt{\alpha'}}$ denotes the gauge coupling. Remarkably, a non-linear $U(1)$ gauge symmetry is known to be preserved in an one form theory in presence of a constant two form on a $D$-brane. In the past, there were several attempts to approximate a non-linear $U(1)$ gauge dynamics by Dirac-Born-Infeld action coupled to Chern-Simmons on a BPS D-brane \cite{bengt-1,bengt-2}. The BPS brane dynamics is known to describe an extremal black hole which corresponds to a near horizon geometry in a string theory.

\sp
\noindent
In the context, a non-linear $U(1)$ gauge dynamics may also be re-expressed in terms of a two form alone on a $D_4$-brane, which is Poincare dual to the one form gauge theory. The duality allows one to address the one form non-linear gauge dynamics, in presence of a constant two form, to a dynamical two form on a $D_4$-brane. The alternate gauge theoretic description by a two form incorporates a non-linear charge and hence leads to a notion of an effective $D_4$-brane. In addition, a Poincare duality interchanges the metric signature between the original and the dual. The two form gauge theory on a $D_4$-brane may be given by
\bea
S_{\rm B}=- {1\over{12C_2^2}}\int d^5x\ {\sqrt{-g}}\ H^2\ ,\label{gauge-3}
\eea
where $C_2^2=(8\pi^3g_s){\alpha'}^{3/2}$ denotes a gauge coupling. The field strength ${H_{\mu\nu}}^{\lambda}$ is governed by the equations of motion of a two form, in presence of a flat background metric $g_{\mu\nu}$, in a $U(1)$ gauge theory. Explicitly, the $B_2$ field equations of motion are given by
\be
\partial_{\lambda}H^{\lambda\mu\nu} + {1\over2} g^{\alpha\beta}\partial_{\lambda}\ g_{\alpha\beta}\ 
H^{\lambda\mu\nu}=0\ . \label{gauge-KerrD-1}
\ee

%%%%%%%%%%%%%%%%%%%%%%%%%%%%%%%%%%%%%%%%%%%%%
\subsection{A non-perturbative setup: Generalized space-time curvature}
%%%%%%%%%%%%%%%%%%%%%%%%%%%%%%%%%%%%%%%%%%%%%
The local degrees in two form, on a $D_4$-brane, have been exploited to construct an effective space-time curvature scalar ${\cal K}^{(5)}$ in a second order formalism \cite{spsk}-\cite{kpss1}. The generalized curvature is primarily sourced by a two form gauge theory on a $D_4$-brane. It has been shown that the space-time effective curvature description becomes essential with a geometric torsion. 
Generically an irreducible scalar ${\cal K}^{(5)}$ governs a geometric torsion ${\cal H}_{\mu\nu\lambda}$, which is primarily described by a gauge theoretic torsion $H_{\mu\nu\lambda}=3\nabla_{[\mu}B_{\nu\lambda ]}$ on a $D_4$-brane. Unlike the extremal brane geometries, the geometrodynamics of a torsion on an effective $D_4$-brane addresses some of the non-extremal quantum vacua in string theory. In fact, the emergent black holes are described by a pair of brane and anti-brane separated by a transverse small dimension in the framework. The $(D{\bar D})_4$-pair breaks the supersymmetry and may describe a non-BPS brane in string theory. Most importantly, an emergent quantum black hole in a low energy limit may seen to describe a classical vacuum in Einstein gravity.

\sp
\noindent
A priori the required modification, to incorporate a geometric notion on a $D_4$-brane, may be viewed through a modified covariant derivative defined with a completely antisymmetric gauge connection: ${H_{\mu\nu}}^{\lambda}$. The appropriate covariant derivative may be given by
\be
{\cal D}_{\lambda}B_{\mu\nu}=\nabla_{\lambda}B_{\mu\nu} + {1\over2}{{H}_{\lambda\mu}}^{\rho}B_{\rho\nu} -{1\over2} {{H}_{\lambda\nu}}^{\rho}B_{\rho\mu}\ .\label{gauge-5}
\ee
Under an iteration $H_3\rightarrow {\cal H}_3$, the geometric torsion in a second order formalism may be defined with all order corrections in $B_2$ in a gauge theory. Formally, a geometric torsion may be expressed in terms of gauge theoretic torsion and its coupling to two form. It is given by
\bea
{\cal H}_{\mu\nu\lambda}&=&3{\cal D}_{[\mu}B_{\nu\lambda ]}\nonumber\\
&=&3\nabla_{[\mu}B_{\nu\lambda ]} + 3{{\cal H}_{[\mu\nu}}^{\alpha}
{B^{\beta}}_{\lambda ]}\ g_{\alpha\beta}
\nonumber\\ 
&=&H_{\mu\nu\lambda} + \left ( H_{\mu\nu\alpha}{B^{\alpha}}_{\lambda} + \rm{cyclic\; in\;} \mu,\nu,\lambda \right )\ +\ H_{\mu\nu\beta} {B^{\beta}}_{\alpha} {B^{\alpha}}_{\lambda} + \dots \;\ .\label{gauge-6}
\eea 
An exact covariant derivative in a perturbative gauge theory may seen to define a non-perturbative covariant derivative in a second order formalism. Thus, a geometric torsion, constructed through a non-perturbative covariant derivative (\ref{gauge-5}), may equivalently be described by an appropriate curvature tensor ${\cal K}_{\mu\nu\lambda\rho}$ which is worked out in ref.\cite{spsk}. Explicitly,
\bea
{{\cal K}_{\mu\nu\lambda}}^{\rho}&=&{1\over2}\partial_{\mu}{{\cal H}_{\nu\lambda}}^{\rho} -{1\over2}\partial_{\nu} {{\cal H}_{\mu\lambda}}^{\rho} 
+ {1\over4}{{\cal H}_{\mu\lambda}}^{\sigma}{{\cal H}_{\nu\sigma}}^{\rho}-{1\over4}{{\cal H}_{\nu\lambda}}^{\sigma}{{\cal H}_{\mu\sigma}}^{\rho}\ ,\nonumber\\ 
{\cal K}_{\mu\nu}&=& -\left (2\partial_{\lambda}{{\cal H}^{\lambda}}_{\mu\nu} +
{{\cal H}_{\mu\rho}}^{\lambda}{{\cal H}_{\lambda\nu}}^{\rho}\right )
\nonumber\\
{\rm and}\qquad {\cal K}&=& -{1\over{4}}{\cal H}_{\mu\nu\lambda}{\cal H}^{\mu\nu\lambda}
\ .\label{gauge-7}
\eea
The fourth order tensor is antisymmetric within a pair of indices, $i.e.\ \mu\leftrightarrow\nu$ and $\lambda\leftrightarrow\rho$, which retains a property of Riemann tensor ${R_{\mu\nu\lambda}}^{\rho}$. However the effective curvature ${{\cal K}_{\mu\nu\lambda}}^{\rho}$ do not satisfy the symmetric property, under an interchange of a pair of indices, as in Riemann tensor. Nevertheless, for a constant torsion the 
generic tensor: ${{\cal K}_{\mu\nu\lambda}}^{\rho}\rightarrow {R_{\mu\nu\lambda}}^{\rho}$. As a result, the effective curvature constructed in a non-perturbative framework may be viewed as a generalized curvature tensor. It describes the propagation of a geometric torsion in a second order formalism.
%%%%%%%%%%%%%%%%%%%%%%%%%%%%%%%%%%%%%%%%%
\subsection{Emergent metric fluctuations}
%%%%%%%%%%%%%%%%%%%%%%%%%%%%%%%%%%%%%%%%%
A geometric torsion ${\cal H}_3$ in a second order formalism may seen to break the $U(1)$ gauge invariance of a two form in the underlying gauge theory.  Nevertheless an emergent notion of metric fluctuation, sourced by a two form local degrees, restores gauge invariance in a generalized irreducible space-time curvature ${\cal K}^{(5)}$. The non-perturbative fluctuations, underlying a $U(1)$ gauge invariance, turn out to be governed by the fluxes and are given by
\be
f_{\mu\nu}^{nz}= C\ {\cal H}_{\mu\alpha\beta}\ {{\bar{\cal H}}^{\alpha\beta}{}}_{\nu}\ 
\approx C\ H_{\mu\alpha\beta}\ {{{\bar{\cal H}}}^{\alpha\beta}{}}_{\nu}\ ,\label{gauge-7}
\ee
where $C$ is an arbitrary constant and ${\bar{\cal H}}_{\mu\nu\lambda}= (2\pi\alpha'){\cal H}_{\mu\nu\lambda}$. The generalized curvature tensor may also be viewed though a geometric field strength ${\cal F}_2$ which is Poincare dual to ${\cal H}_3$ on a $D_4$-brane. Then, a geometric ${\cal F}_2$ may be given by
\bea
{\cal F}_{\alpha\beta}&=&{\cal D}_{\alpha}A_{\beta} -{\cal D}_{\beta}A_{\alpha}\nonumber\\
&=&\left ( {\cal F}^z_{\alpha\beta} + {{\cal H}_{\alpha\beta}}^{\delta}A_{\delta}\right )\ ,\label{gauge-8}
\eea
where ${\bar{\cal F}}^z_{\alpha\beta}= (2\pi\alpha'){\cal F}^z_{\alpha\beta}=\left ({\bar F}_{\alpha\beta}+ B^z_{\alpha\beta}\right )$ is defined with a zero mode $B^z_{\mu\nu}$ on a $D_4$-brane. It signifies a non-linear electromagnetic field and is gauge invariant under a non-linear $U(1)$ transformations \cite{seiberg-witten}. Apparently a non-zero $H_3$ seems to break the $U(1)$ gauge invariance. Nevertheless, an action defined with a lorentz scalar ${\cal F}^2$ may seen to retain the gauge invariance with the help of an emerging notion of metric fluctuations in the formalism. Then, the fluctuations (\ref{gauge-7}) in its dual description may be given by
\be
f_{\mu\nu}^{nz}={\tilde C} {\bar{\cal F}}_{\mu\alpha}{{\bar{\cal F}}^{\alpha}{}}_{\nu}\ ,\label{gauge-9}
\ee
where ${\tilde C}$ is an arbitrary constant. The dynamical fluctuations in eqs.(\ref{gauge-7}) and (\ref{gauge-9}) modify the constant metric
on a $D_4$-brane. Then, the emergent metric on a $D_4$-brane becomes 
\bea
G_{\mu\nu}&=&\left ( G^z_{\mu\nu}\ +\ C\ {\bar{\cal H}}_{\mu\lambda\rho}\ {{\cal H}^{\lambda\rho}{}}_{\nu}\right )\nonumber\\
&=&\left ( G^z_{\mu\nu}\ +\ {\tilde C}\ {\bar{\cal F}}_{\mu\lambda}{{\bar{\cal F}}^{\lambda}{}}_{\nu}\right )
\ .\label{gauge-10}
\eea 
The fluxes, in a bilinear combination are gauge invariant and hence an emergent metric a priori seems to be unique. However an analysis reveals that 
the emergent metric may not be unique due to the coupling of $B_2$-potential to $H_3$ underlying a geometric torsion ${\cal H}_3$. In other words, $B_2$-fluctuations do play a significant role to define the emergent geometries in a non-perturbative framework. They lead to a generalized emergent metric on a $D$-brane. It may be given by
\be
G_{\mu\nu}=\left ( g_{\mu\nu}\ -  B_{\mu\lambda}{B^{\lambda}}_{\nu}\ +\ C\ {\bar{\cal H}}_{\mu\lambda\rho}\ {{\cal H}^{\lambda\rho}{}}_{\nu}
+\ {\tilde C}\ {\bar{\cal F}}_{\mu\lambda}{{\bar{\cal F}}^{\lambda}{}}_{\nu}\right )
\ .\label{gauge-11}
\ee
The background fluctuations, arising out of the non-dynamical components in $B_{\mu\nu}$, may seen to deform the brane geometries significantly. They may lead to a large number of vacua and may correspond to the landscape quantum geometries in the formalism. The background fluctuations in two form 
may have their origin in a higher dimensional gauge theoretic torsion $H_3$. They may couple to an electro-magnetic field in higher dimensions to define a gauge invariant non-linear ${\cal F}^z_2$. The gauge dynamics on a $D_4$-brane, in presence of gauge connections, may be approximated by an irreducible generalized curvature theory in a second order formalism. A priori, the effective curvature may seen to describe a geometric torsion dynamics on an effective $D_4$-brane \cite{spsk,spsk-RN}. A geometric construction of a torsion in a non-perturbative framework is inspiring and may provoke thought to unfold certain aspects of quantum gravity. Generically, the action may be given by
\be
S_{\rm D_4}^{\rm eff}= {1\over{3C_4^2}}\int d^5x {\sqrt{-G}}\ \left ( {\cal K}^{(5)}- {\Lambda}\right )\ ,\label{gauge-12}
\ee
where $C_4^2=(8\pi^3g_s){\alpha'}^{3/2}$ is a constant and $G=\det G_{\mu\nu}$. The cosmological constant ${\Lambda}$, in the geometric action, is sourced by a zero mode in the theory. The covariant derivative satisfies ${\cal D}_{\lambda}{G}_{\mu\nu}=0$. Thus, an emergent metric in the framework uniquely fixes the covariant derivative. On the other hand, the effective curvature theory may formally be viewed as a non-linear $U(1)$ gauge theory. The
energy-momentum tensor may be given by
\bea
T_{\mu\nu}&=&{1\over{6}}\left ({{\Lambda}-{\cal K}^{(5)}}\right ){G}_{\mu\nu} - {1\over{8C\pi\alpha'}}f_{\mu\nu}^{nz}\nonumber\\
&=&{1\over6}\left ( {{\Lambda}-{\cal K}^{(5)}}\right )G^z_{\mu\nu} + \left ( {{\Lambda-{\cal K}^{(5)}}\over{6}} - {1\over{8C\pi\alpha'}}\right )f_{\mu\nu}^{nz}\ . \label{gauge-13}
\eea
The trace of energy-momentum tensor on a $D_4$-brane becomes
\be
T={{{\cal K}^{(5)}}\over6} + {{5\Lambda}\over6}\ .\label{gauge-14}
\ee
It ensures that a vacuum, $i.e.\ T=0$, may be defined in a gauge choice for ${\cal K}^{(5)}=0$ and $\Lambda=0$.
With a gauge ${\Lambda}=(3/{\pi\alpha'})+{\cal K}$, the $T_{\mu\nu}$ may be viewed to source an emergent metric in a non-perturbative framework. It is given by
\bea
T_{\mu\nu}&=&\left ( {{G^z_{\mu\nu}}\over{2\pi\alpha'}}\ +\ \left [ C-{1\over4}\right ]\ {\cal H}_{\mu\lambda\rho} {{\cal H}^{\lambda\rho}}_{\nu}\right )\nonumber\\
&=&\left ( {{G_{\mu\nu}}\over{2\pi\alpha'}}\ - {1\over4} {\cal H}_{\mu\lambda\rho} {{\cal H}^{\lambda\rho}}_{\nu}\right )\ .\label{gauge-31}
\eea
Thus, the $T_{\mu\nu}$ in a gauge theory sources the geometro-dynamics of a torsion in a generalized curvature theory. A higher dimensional $T_{\mu\nu}$ can source a lower dimensional background fluctuations in two form on a brane. 

%%%%%%%%%%%%%%%%%%%%%%%%%%%%%%%%%%%%%%%%%%%%%%%%%%%%%%%%%%%%%%%%%%%%%%
\section{Kerr Geometries on ${\mathbf{(D{\bar D})_4}}$-brane}
%%%%%%%%%%%%%%%%%%%%%%%%%%%%%%%%%%%%%%%%%%%%%%%%%%%%%%%%%%%%%%%%%%%%%%
In this section, we focus on the five dimensional aspects of a generalized curvature underlying a non-linear $U(1)$ gauge theory on a $D_4$-brane. 
We use a gauge choice for a two form to construct a quantum Kerr black hole in an emergent gravity scenario. 

\sp
\noindent
A priori, an effective $D_4$-brane may seen to be influenced by a generic torsion in the formalism. Nevertheless, we argue that the torsion completely decouples to yield a stable vacuum both in the semi-classical and in the quantum regimes. Formally, the equations of motion (\ref{gauge-KerrD-1}) may alternately be viewed through ${\cal D}_{\lambda}{\cal H}^{\lambda\mu\nu}=0$. However, the gauge invariance in presence of a geometric torsion 
enforces a space-time generalized curvature in the framework. In the paper, we consider a flat background metric in Boyer-Lindquist coordinates. It is described by a Minkowski vacuum on a $D_4$-brane. The line-element is given by
\be
ds^2=- dt^2 +{{\rho^2}\over{\tri}}\ dr^2+\rho^2\ d\t^2 +(r^2+a^2)\sin^2\t \ d\phi^2 +(r^2+b^2) \cos^2\t\ d\psi^2\ ,\label{kerr-brane-3}
\ee
\vspace{-.2in}
$${\rm where}\;\ \rho^2=\left ( r^2 +a^2\cos^2\t +b^2\sin^2\t \right )\;\; {\rm and}\;\ \tri=\left ( r^2 + a^2)(r^2 +  b^2\right )\ .\qquad\qquad {}$$
The effective radius at the pole(s) and on the equator are, respectively, denoted by $\rho_{\rm p}={\sqrt{r^2+a^2}}$ and $\rho_{\rm e}={\sqrt{r^2+b^2}}$. The range of the angular coordinates are: ($0<$$\phi$$<$$2\pi ,\ 0$$<$$\psi$$<$$2\pi ,\ 0$$<$$\theta$$<$$\pi$). They may be expressed by the cartesian coordinates:
\bea
&&x=\rho_{\rm p}\ \sin\t \cos\phi\ ,\quad\qquad\qquad\; y=\rho_{\rm p}\ \sin\t  \sin\phi\; ,\nonumber\\
&&z=\rho_{\rm e}\ \cos\t \cos\psi\qquad\; {\rm and} \qquad w=\rho_{\rm e}\ \cos\t \sin\psi\ .\label{kerr-brane-2}
\eea
The coodinates ensure a circle in $xy$-plane for $\t\neq (0,\pi )$ and in $wz$-plane for $\t\neq \pi/2$. The equations are:
$$
(x^2+y^2)\ =\ \rho_{\rm p}^2\ \sin^2\t \qquad {\rm and}\quad (z^2+w^2)\ =\ \rho_{\rm e}^2\ \cos^2\t\ .
$$
Thus the Minkowski line element is chracterized by an underlying ring, which orient itself from a $wz$-plane at the poles to a $xy$-plane on the equator.
%%%%%%%%%%%%%%%%%%%%%%%%%%%%%%
\subsection{Two form ansatz-I}
%%%%%%%%%%%%%%%%%%%%%%%%%%%%%%
An emergent Kerr vacuum in five dimensions, leading to a Kerr black hole in Einstein gravity, enforces a non-propagating torsion in the generalized curvature theory on a $(D{\bar D})_4$-brane. A torsion free gravity ensures a vacuum $T_{\mu\nu}=0$ in the framework. In the context, a gauge choice for
a two form may be worked out to yield: 
\bea
&&B_{tr}\ =\ r\sqrt{2M\over{\tri}}\; ,\qquad\qquad\qquad\qquad\qquad\qquad\qquad {}\nonumber\\
&&B_{\t\psi}\ =\ b{\sqrt{2M}}\ \cos^2\t\; ,\nonumber\\
&&B_{\t\phi}\ =\ a{\sqrt{2M}}\ \sin^2\t\; ,\nonumber\\
&&B_{r\t}\ =\ \rho\left (2 +{{(2M-\rho^2)r^2}\over{\tri}} + {{2Mr^2+\tri}\over{r^2\rho^2}}\right)^{1/2}\nonumber\\
&&\quad\;\;\;\; =\ \rho \left ( {{{\cal M}^+}\over{\tri}}\  -\ {{{\cal M}^-}\over{r^2\rho^2}}\right )^{1/2}\ ,\label{kerr-brane-1}
\eea
\vspace{-.1in}
$${\rm where}\qquad {\cal M}^{\pm}\ =\ \left ( a^2\sin^2\t + b^2 \cos^2\t \pm 2M\right )r^2\ +\ a^2b^2\ .\qquad\qquad\qquad\qquad {}$$
The $B_{\mu\nu}$ potential is assumed to be dimensionless and we set $(2\pi\alpha')=1$ throughout this paper. The constants $(a,b,M)>0$ are arbitrary and they may be identified with the conserved quantities defined in an asymptotic regime underlying an effective geometry on a brane. 

\sp
\noindent
The gauge choice re-assures a vanishing torsion $H_3=0={\cal H}_3$ and ${\cal F}_2\rightarrow {\cal F}^z_2\rightarrow F_2$, in presence of the $B_2$-fluctuations on a $D_4$-brane. Nevertheless, a non-propagating geometric torsion, in a gauge choice on a $D_4$-brane, may have its origin in a higher dimensional $D$-brane. Thus, a non-propagating torsion on a $D_4$-brane may possess local degrees in a higher dimensional brane. On the other hand, a linear $F_2$ may be gauged away to signify that an emergent geometry on a $D_4$-brane is a priori sourced by the background fluctuations in $B_2$. They re-assure a generalized nature of the irreducible curvature scalar ${\cal K}^{(5)}$ underlying a geometric torsion in a non-perturbative farmework.

%%%%%%%%%%%%%%%%%%%%%%%%%%%%%%%%%%%%%%%%%%%%%%
\subsection{Quantum Kerr black hole: ansatz-I}
%%%%%%%%%%%%%%%%%%%%%%%%%%%%%%%%%%%%%%%%%%%%%%
In a gauge choice the generic metric (\ref{gauge-11}) underlying a generalized space-time curvature theory on an effective $D_4$-brane reduces 
drastically. It takes a form:
\be
G_{\mu\nu}\rightarrow\left ( g_{\mu\nu}\ -\ B_{\mu\lambda}g^{\lambda\rho}B_{\rho\nu}\right )\ .\label{kerr-brane-4}
\ee
Explicitly the emergent metric components, on a brane, are worked out to yield:
\bea
G_{tt}&=&-\left (1-\frac{2M}{\rho^2}\right )\ ,\nonumber\\ 
G_{rr}&=&\left (1- {{{\cal M}^-}\over{r^2\rho^2}}\right )\ ,\nonumber\\
G_{\t\t}&=&2M\left ( 2- {{r^2\rho^2}\over{\tri}}\right )\ + \ {{\tri}\over{r^2}}\ G_{rr}\nonumber\\
&=&\rho^2\ + \left [ {{{\cal M}^+}\over{r^2}} + {{\tri}\over{r^2}} \left (G_{rr}-1\right )
 + {{2Mr^2}\over{\tri}}\left ({{{\cal M}^+}\over{r^2}} - 2M\right )\right ]\nonumber\\
&=&\rho^2\ +\ \rho^2_0\ ,\nonumber\\
G_{\phi\phi}&=&\left (r^2+a^2+\frac{2a^2M\sin^2\t}{\rho^2}\right )\sin^2\t\ ,\nonumber\\ 
G_{\psi\psi}&=&\left (r^2+b^2+\frac{2b^2M\cos^2\t}{\rho^2}\right )\cos^2\t\ ,\nonumber\\
G_{t\t}&=&-{{\sqrt{2M\tri}}\over{r\rho}}\Big ({{{\cal M}^+}\over{\tri}} + G_{rr}-1\Big )^{1/2}\ ,\nonumber\\
G_{r\phi}&=&-{a{\sqrt{2M}}\over{\rho}}\left ({{{\cal M}^+}\over{\tri}} +G_{rr}-1\right )^{1/2}\sin^2\t\ ,\nonumber\\
G_{r\psi}&=&-{{b{\sqrt{2M}}}\over{\rho}}\left ({{{\cal M}^+}\over{\tri}}+G_{rr}-1\right )^{1/2}\cos^2\t\ ,\nonumber\\
G_{\phi\psi}&=&{{2abM}\over{\rho^2}}\ \sin^2\t \cos^2\t\ .\label{kerr-brane-5}
\eea
For a calculational simplicity, we restrict the emergent quantum geometry to a window defined by 
$$\rho^4>>(M^2,a^4,b^4)\ ,\;\; {\rm with}\;\ \rho^2>(M,a^2,b^2)\ ,\;\; {\rm for\ a\ fixed}\;\ \alpha'\ .$$ 
The limits imply $r^4$$>$$>$$(M^2,a^4,b^4)$ with $r^2$$>$$(M,a^2,b^2)$, which naturally forbid $r\rightarrow 0$ on a brane geometry. 
The brane window truncates the ultra high energetic geometric modes in the full quantum gravity. In the regime, the emergent metric component $G_{rr}$ may be approximated to yield:
\be
G_{rr}\rightarrow \left (1- {{2M}\over{\rho^2}} + {{\left (a^2\sin^2\t + b^2 \cos^2\t \right )r^2 + a^2b^2}\over{r^2\rho^2}}\right )^{-1}\ .\label{kerr-brane-6}
\ee
In the limit: $(G_{rr}-1)\approx (1-G_{rr}^{-1})$. 
The emergent causal patches in the regime on a brane may imply a rotating black hole characterized by an event horizon and an ergo-sphere. The horizon radii $r_{\pm}$ are computed from $G_{rr}^{-1}=0$. A priori, they are given by
\bea
r_{\pm}&=&{1\over{\sqrt{2}}}\left (2M-a^2-b^2\pm \sqrt{(2M-a^2-b^2)^2-4a^2b^2} \right )^{1/2}\nonumber\\
&=& {\sqrt{M}}\left ( 1-\left ({{a^2+b^2}\over{2M}}\right )\pm {\sqrt{1+ \left ({{a^2-b^2}\over{2M}}\right )^2 -\left ({{a^2+b^2}\over{M}}\right )}}\right )^{1/2}\ .\label{kerr-brane-6}
\eea
At the horizons, the effective radii are: $\rho_{\pm}= \left (r_{\pm}^2 + a^2\cos^2\t_{\pm} + b^2 \sin^2\t_{\pm}\right )^{1/2}$.
The geometric patches (\ref{kerr-brane-5}) enforce a lower bound on its mass $i.e.\ M\ge (a^2+b^2)$ for $a\neq b$. For $a=b$, the black hole mass satisfies an inequality $M>(a^2+b^2)$. Generically for all $a$ and $b$, the lower bound is ${\sqrt{M}}\ge (a+b)$. The torsion dynamics leads to a spinning black hole defined with an ergo sphere. The ergo radius is computed from $G_{tt}=0$ and we obtain
\bea
r_{\rm ergo}&=&{\sqrt{2M-a^2\cos^2\t -b^2\sin^2\t}} \nonumber\\
&=&{\sqrt{2M-a^2}}\left ( 1 + {{a^2-b^2}\over{2M-a^2}}\ \sin^2\t \right )^{1/2}\ .\label{kerr-brane-7}
\eea
For $a\neq b$, the ergo radius varies with polar angle from ${\sqrt{2M-a^2}}$ at poles to ${\sqrt{2M-b^2}}$ on the equator. When $a=b$, the ergo radius turns out to take a fixed value ${\sqrt{2M-a^2}}$. 
\begin{figure}
\mbox{
\subfigure{\includegraphics[width=0.45\textwidth,height=0.21\textheight]{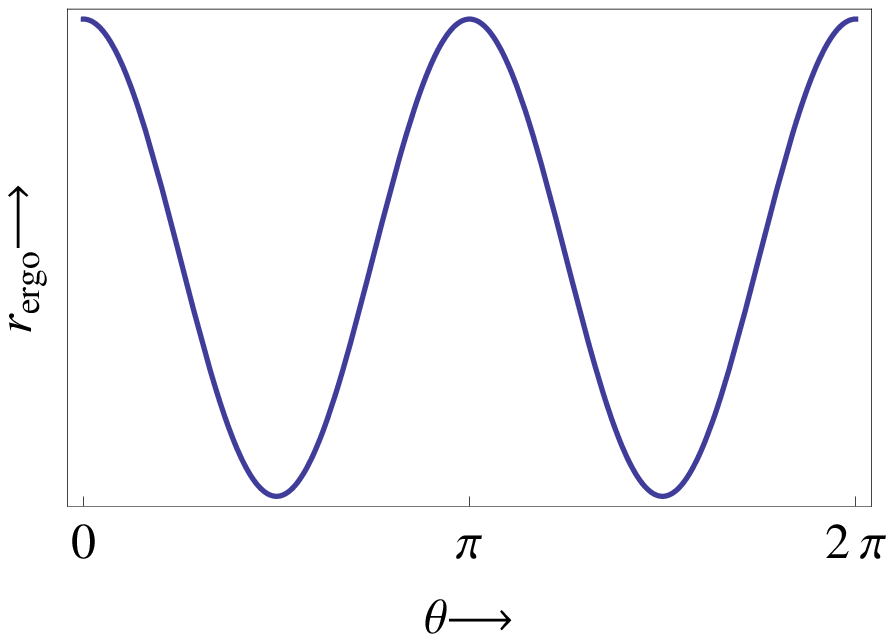}}
\hspace{.25in}
\subfigure{\includegraphics[width=0.45\textwidth,height=0.21\textheight]{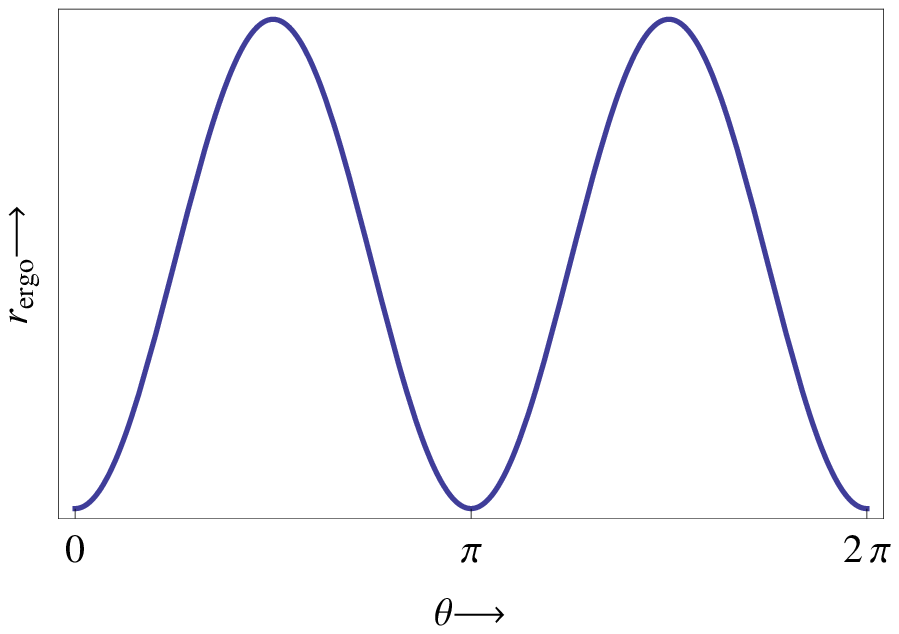}}}
%\mbox{\includegraphics[width=0.53\linewidth,height=0.31\textheight]{Horizon1}}
%\vspace{-.5in}
\caption{\it Variations of $r_{\rm ergo}$ with the polar angle $\t$, for (i) $b>a$ and (ii) for $a>b$, in a $5D$ Kerr black hole. They differ by a phase of $\pi/2$. A generic ergo sphere with a varying radius for $a\neq b$, or a fixed radius for $a=b$, covers the event horizon. However for certain polar angles, a new conserved charge may seen to describe the Kerr geometry, which may be viewed at the expense of a vanishing angular momentum.}
\end{figure}
\noindent
Now we perform an interchange $dt\leftrightarrow dr$ in the off-diagonal metric components to tune the quantum geometry with the appropriate conserved charges in a low energy limit. The interchange, flips: 
(two charges and $\Omega^{\t}$) $\leftrightarrow$ ($\Omega^{\phi}$, $\Omega^{\psi}$, and a charge). However, the causal quantum patches remain unchanged under the interchange and they ensure the characteristics of a quantum Kerr black hole. Under $dt\leftrightarrow dr$ in the off-diagonal metric components, the emergent geometry becomes
\bea
ds^2&=&-\ \left (1-\frac{2M}{\rho^2}\right )dt^2\ +\ \left (1- {{2M}\over{\rho^2}} + {{\left ({a^2\sin^2\t + b^2 \cos^2\t}\right )r^2+  {a^2b^2}}\over{r^2\rho^2}}\right )^{-1} dr^2\ +\  \left ( \rho^2 + {\rho}^2_0\right ) d\t^2\nonumber\\
&&+\ \left (r^2+a^2+\frac{2a^2M\sin^2\t}{\rho^2}\right )\sin^2\t\ d\phi^2\ +\ \left (r^2+b^2+\frac{2b^2M\cos^2\t}{\rho^2}\right )\cos^2\t\ d\psi^2\nonumber\\ 
&&+\ {{4abM\ {\sin^2\t \cos^2\t}}\over{\rho^2}} d\phi d\psi\ -\ {2{\sqrt{2M}}\over{\rho}}\left (4M + \left (1-G^{-1}_{rr}\right )\left [ {{\tri}\over{r^2}} - \rho^2\right ]\right )^{1/2}drd\t\ \nonumber\\
&&-\ {2a{\sqrt{2M}}\over{\rho}}\left ({{4Mr^2}\over{\tri}} + \left (1-G_{rr}^{-1}\right )\left [1 - {{r^2\rho^2}\over{\tri}}\right ]\right )^{1/2}\sin^2\t\ dt d\phi\ \nonumber\\
&&-\ {2b{\sqrt{2M}}\over{\rho}}\left ({{4Mr^2}\over{\tri}} + \left (1-G_{rr}^{-1}\right )\left [1 - {{r^2\rho^2}\over{\tri}}\right ]\right )^{1/2}\cos^2\t\ dt d\psi\ .\label{kerr-brane-8}
\eea
A priori, the emergent geometry describes a quantum Kerr black hole in five dimensions. Under $r\rightarrow -r$, an emergent Kerr black hole on a $D_4$-brane transforms to that on an anti $D_4$-brane. The ${\bar D}_4$- and $D_4$- propagate, respectively, along $-r$ and $+r$ in the near horizon. They move along the opposite time-like directions connecting the infinite past to the infinite future within the event horizon. Their geometries would differ only by a sign in the $G_{r\t}$ component. In a global scenario, $i.e.$ a pair of $(D{\bar D})_4$-brane, the charge due to the $G_{r\t}$ may seen to nullify in a quantum Kerr black hole. Then the appropriate, non-extremal, quantum geometry on a pair of $(D{\bar D})_4$-brane is given by
\bea
ds^2&=&-\ \left (1-\frac{2M}{\rho^2}\right )dt^2\ +\ \left (1- {{2M}\over{\rho^2}} + {{\left ({a^2\sin^2\t + b^2 \cos^2\t}\right )r^2+  {a^2b^2}}\over{r^2\rho^2}}\right )^{-1} dr^2\ +\  \left ( \rho^2 + {\rho}^2_0\right ) d\t^2\nonumber\\
&&+\ \left (r^2+a^2+\frac{2a^2M\sin^2\t}{\rho^2}\right )\sin^2\t\ d\phi^2\ +\ \left (r^2+b^2+\frac{2b^2M\cos^2\t}{\rho^2}\right )\cos^2\t\ d\psi^2\nonumber\\ 
&&-\ {2a{\sqrt{2M}}\over{\rho}}\left ({{4Mr^2}\over{\tri}} + \left (1-G_{rr}^{-1}\right )\left [1 - {{r^2\rho^2}\over{\tri}}\right ]\right )^{1/2}\sin^2\t\ dt d\phi\ +\ {{4abM\ {\sin^2\t \cos^2\t}}\over{\rho^2}} d\phi d\psi\nonumber\\
&&-\ {2b{\sqrt{2M}}\over{\rho}}\left ({{4Mr^2}\over{\tri}} + \left (1-G_{rr}^{-1}\right )\left [1 - {{r^2\rho^2}\over{\tri}}\right ]\right )^{1/2}\cos^2\t\ dt d\psi\ .\label{kerr-brane-9}
\eea
\begin{figure}
\mbox{
\subfigure{\includegraphics[width=0.45\textwidth,height=0.21\textheight]{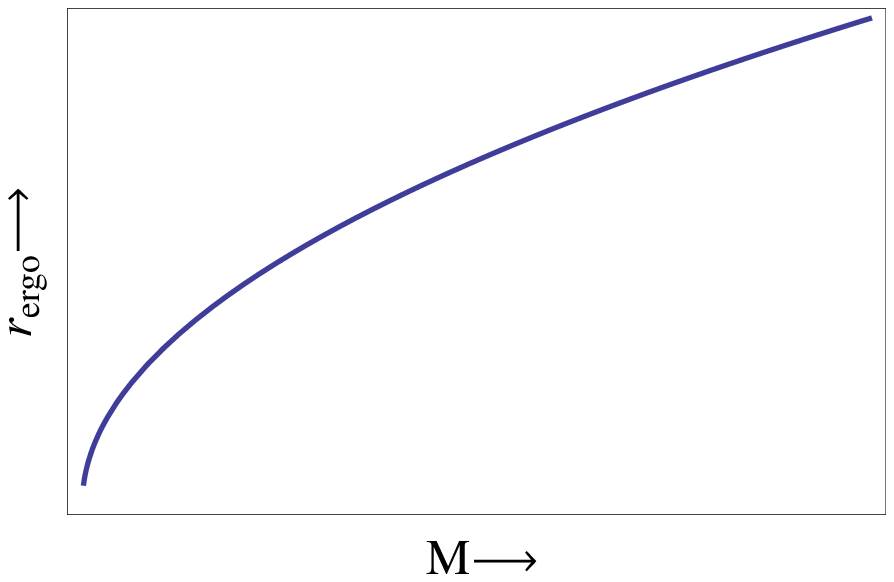}}
\hspace{.15in}
\subfigure{\includegraphics[width=0.45\textwidth,height=0.21\textheight]{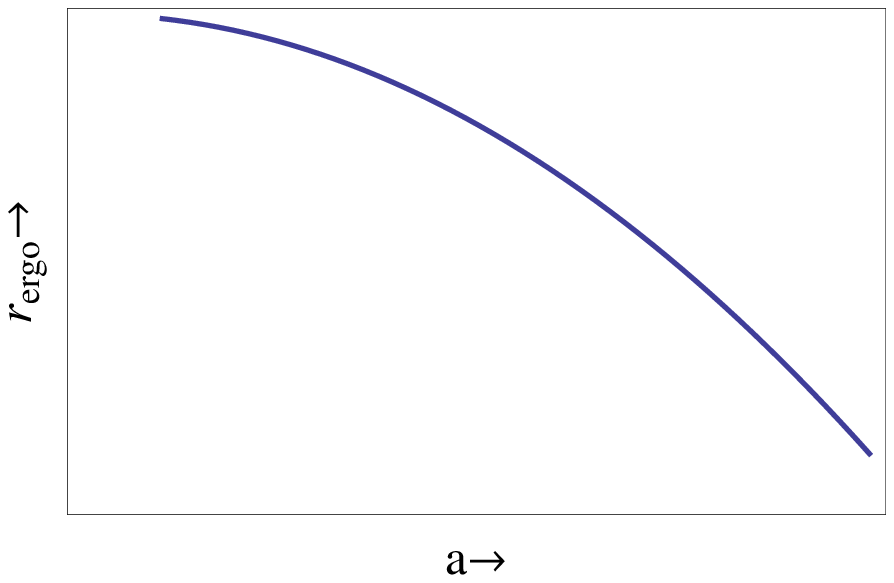}}}
%\mbox{\includegraphics[width=0.53\linewidth,height=0.31\textheight]{Horizon1}}
%\vspace{-.2in}
\caption{\it The ergo radius $r_{\rm ergo}$ increases with an increase in black hole mass $M$ for fixed ($a,b$) in a 5D Kerr black hole. However, the ergo sphere apparently shrinks with an increased value of the background parameter: $a$ for fixed ($M,b$) or $b$ for a fixed ($M,a$) and vice-versa. It implies that a macroscopic black hole may also be described in a limit of an extremely small $a$- or $b$-perturbations to the $S^3$-symmetric vacuum on a $D_4$-brane.}
\end{figure}
%\vspace{-.2in}
\noindent
A non-zero ergo radius (\ref{kerr-brane-7}) incorporates an angular velocity into the brane black hole. It is computed at the event horizon to yield:
\bea
\Omega^{\phi}&=&-\left( {{a\rho_+}\over{r_+}}\right ) {{\sqrt{\left (a^2\sin^2\t_+ +b^2 \cos^2\t_+ + 4M\right )r_+^2 + a^2b^2}}\over{\rho_+^2\left (r_+^2+a^2\right ) + 2Ma^2 \sin^2\t_+ }}\nonumber\\
&\approx& -a{\sqrt{a^2 +b^2 + 4M}\over{r_+^3}}\nonumber\\ \nonumber\\ 
{\rm and}\quad \Omega^{\psi}&=&-\left ({{b\rho_+}\over{r_+}}\right ) 
{{\sqrt{\left (a^2\sin^2\t_+ +b^2 \cos^2\t_+ + 4M\right )r_+^2 + a^2b^2}}\over{\rho_+^2\left (r_+^2+b^2\right ) + 2Mb^2 \cos^2\t_+ }}\nonumber\\
&\approx&-b{{\sqrt{a^2 +b^2 + 4M}}\over{r_+^3}}\ .\label{kerr-brane-10}
\eea
It shows that a microscopic black hole spins much faster than a macroscopic black hole. The curvature singularity in $R_{\mu\nu\lambda\rho}R^{\mu\nu\lambda\rho}$ in the quantum Kerr black hole is at $\rho\rightarrow 0$. In the limit, the curvature singularity is 
described by the equation of a circle. Generically they are given by
$$(x^2+y^2)\ =\ (a^2-b^2)\sin^4\t\qquad {\rm and}\quad (z^2+w^2)\ =\  (b^2-a^2)\cos^4\t\ .$$ 
\noindent
They imply a ring singularity on a equatorial plane for $a>b$ and at poles for $a<b$. In a simplified case for $a=b$, the 5D Kerr black hole may seen to possess a point singularity. The curvature singularity at $\rho\rightarrow 0$, in the quantum Kerr black hole, reduces to $\rho_p\rightarrow 0$ on the equator and $\rho_e\rightarrow 0$ at the poles. Since the curvature singularity at $\rho\rightarrow 0$ is not accessible for the real background parameters ($a,b$), the quantum Kerr black hole in 5D may be viewed as a trapped geometry on a non-BPS brane. The forbidden limit $\rho\rightarrow 0$, may imply an intrinsic minimal length scale in the quantum Kerr black hole.

%%%%%%%%%%%%%%%%%%%%%%%%%%%%%%%%%%%%%%%%%%%%%%%%%%%%%%%%%%%%%%%%%%%%%%%%%
\subsection{Emergent Kerr black hole in 5D: ansatz-I}
%%%%%%%%%%%%%%%%%%%%%%%%%%%%%%%%%%%%%%%%%%%%%%%%%%%%%%%%%%%%%%%%%%%%%%%%%
The low energy vacuum in the emergent quantum gravity may be obtained in a large $r$ limit. In particular, the limit is defined by $r^2>>1$ which
truncates a quantum Kerr black hole (\ref{kerr-brane-9}) to a classical Kerr vacuum. With a subtlety, underlying the small $a-$ and $b-$ background parameters, the low energy limit may also be refined by ${{\tri}}\rightarrow (r\rho)^2$. For instance, the $\rho_0^2$ in eq.(\ref{kerr-brane-5}) drastically reduces in the limit to yield:
\bea
\rho_0^2&=&4M\ +\ {{2M}\over{\rho^2}}\left (a^2\sin^2\t + b^2\cos^2\t\ +\ {{a^2b^2}\over{r^2}} \right )\nonumber\\
&=&2M\left ( 1 \ +\ {{\tri}\over{r^2\rho^2}}\right )\ \rightarrow\ 4M\ .\label{kerr-brane-105}
\eea
In the limit, the quantum geometry becomes
\bea
ds^2&=&-\left(1-\frac{2M}{\rho^2}\right) dt^2\ +\ \left (\rho^2+4M\right )\ d\t^2\ +\ \left(1-\frac{2M}{\rho^2}+\frac{\left (a^2\sin^2\t +b^2\cos^2\t\right )r^2 +a^2b^2}{r^2\rho^2} \right)^{-1} dr^2\nonumber\\ 
&&+ \left ( r^2+a^2 + \frac{2a^2M\sin^2\t}{\rho^2}\right) \sin^2\t \ d\phi^2\ + \left(r^2+b^2+\frac{2b^2M\cos^2\t}{\rho^2}\right) 
\cos^2\t \ d\psi^2\nonumber\\
&&-\ \frac{(4{\sqrt{2}})aM\sin^2\t}{\rho^2}\ dtd\phi\ -\ \frac{({4{\sqrt{2}}})bM\cos^2\t}{\rho^2}\ dtd\psi\ +\ \frac{4abM\cos^2\t \sin^2\t}{\rho^2}d\phi d\psi\ .\label{kerr-brane-11}
\eea
Thus, the quantum Kerr black hole on a non-BPS brane in the regime precisely corresponds to a typical Kerr black hole established as a 5D vacuum in Einstein gravity. Interestingly, the causal patches in the quantum geometries remain unaffected in a low energy limit. They imply an exact causal patches in a non-linear gauge theory on a $D_4$-brane. Alternately, the result may be viewed as an artifact of an underlying non-pertubative curvature theory. The classical black hole obtained from a quantum Kerr vacuum on a brane may be re-expressed in a familiar form \cite{sakaguchi}. It is given by
\bea
ds^2&=&-\ dt^2\ +\ {{\rho^2}\over{\tri}} dr^2\ +\ \rho^2 \left( d\t^2 + \sin^2\t \ d\phi^2 + \cos^2\t\ d\psi^2\right ) + {{2M}\over{\rho^2}} dt^2 \nonumber\\
&&+\ \left (a^2-b^2 + {{2Ma^2}\over{\rho^2}}\right ) \sin^4\t\ d\phi^2 + \left (b^2-a^2 + {{2Mb^2}\over{\rho^2}}\right )\cos^4\t\ d\psi^2\nonumber\\
&&-\ {{(4{\sqrt{2}})aM\sin^2\t}\over{\rho^2}}\ dtd\phi -{{{(4\sqrt{2})}bM\cos^2\t}\over{\rho^2}}\ dtd\psi\ +\ {{4abM \cos^2\t \sin^2\t}\over{\rho^2}}\ d\phi d\psi\ .\label{kerr-brane-111}
\eea
For $a=b$, the deformations are simplified. It further re-confirms a generalized nature of ${\cal K}^{(5)}$, which reduces to a Ricci scalar in a gauge choice (\ref{kerr-brane-1}). It is remarkable to note that the $B_2$-fluctuations, on an effective $D_4$-brane, may describe a quantum Kerr black hole with two angular momenta in five dimensions. The vanishing energy-momentum tensor further re-assures a vacuum solution.
%Nevertheless, a $B_2$-fluctuations may have their origin in a higher ($p$$>$$4$) dimensional $D_p$-brane describing a dynamical torsion. For instance, a system of a $D_4$-brane and ${\bar D}_4$-brane cancels the Ramond-Ramond (RR) charge of each other in type II superstring on $S^1$ to nucleate a $D_5$-brane \cite{spsk}. In principle, the nucleation of a higher dimensional $D_p$-brane from its lower dimensional $(D{\bar D})_{(p-1)}$ system saturates at $p=8$ in the formalism.

%%%%%%%%%%%%%%%%%%%%%%%%%%%%%%%%%%%%%%%%%%%%%%%%%%%%%%%%%%%%%%%%%%%%%%%%%
\section{Kerr tunneling vacua on a ${\mathbf{(D{\bar D})_4}}$-brane}
%%%%%%%%%%%%%%%%%%%%%%%%%%%%%%%%%%%%%%%%%%%%%%%%%%%%%%%%%%%%%%%%%%%%%%%%%

%%%%%%%%%%%%%%%%%%%%%%%%%%%%%%%%%%%%%%%%%%%%%%%%%%%%%%%%%%%%%%%%%%%%%%%%%
\subsection{Two form ansatz-II}
%%%%%%%%%%%%%%%%%%%%%%%%%%%%%%%%%%%%%%%%%%%%%%%%%%%%%%%%%%%%%%%%%%%%%%%%%
A two form gauge transformation in a $U(1)$ gauge theory generates an infinite number of nontrivial potentials on a $D_4$-brane. Their coupling to the gauge invariant $H_3$ lead to a large number of emergent quantum vacua (\ref{gauge-11}) in the generalized curvature theory. Then, the gauge choice (\ref{kerr-brane-1}) leading to an emergent Kerr black hole in five dimensions on a brane is not unique. It is plausible to believe that an effective $D$-brane may describe a large number of vacua in a landscape including some of those known in Einstein gravity. Now, we consider a different ansatz for a two form undergoing fluctuations on a $D_4$-brane. For the arbitrary constants ${\tilde a}$ and ${\tilde b}$, the two form ${\tilde B}_2$ ansatz-II may be formally be expressed in terms of the ansatz-I for $B_2$ in eq.(\ref{kerr-brane-1}). They are:
\bea
{\tilde B}_{tr}&=&r{\sqrt{{2M}\over{\tri}}}\;\;\ \longrightarrow\; B_{tr}\ ,\nonumber\\
{\tilde B}_{r\psi}&=&\tilde{b}r{\sqrt{{2M}\over{\tri}}}\;\ \longrightarrow \ {r\over{\sqrt{\tri}}} \left [{{\tilde b}\over{b\cos^2\t}}\right ]\ B_{\t\psi}\ ,\nonumber\\
{\tilde B}_{r\phi}&=&\tilde{a}r{\sqrt{{2M}\over{\tri}}}\; \longrightarrow \ {r\over{\sqrt{\tri}}} \left [{{\tilde a}\over{a\sin^2\t}}\right ]\ B_{\t\phi}\ ,\nonumber\\
{\tilde B}_{r\t}&=&\rho\left ({{{\cal M}^+}\over{\tri}} - {{{\cal M}^-}\over{r^2\rho^2}} -
{{2{\tilde a}^2Mr^2}\over{\tri\left (r^2+a^2\right )\sin^2\t}} - {{2{\tilde b}^2Mr^2}\over{\tri\left (r^2+b^2\right )\cos^2\t}}\right )^{1/2}\nonumber\\
&&\longrightarrow\ \left (B_{r\t}^2 - {{2Mr^2\rho^2}\over{\tri}} \left [ {{{\tilde a}^2}\over{(r^2+a^2)\sin^2\t}} + {{{\tilde b}^2}\over{(r^2+b^2)\cos^2\t}}\right ]\right )^{1/2}.\label{kerr-brane-21}
\eea
In addition to the functional factors, in two of the components, and the corrections in an other component (\ref{kerr-brane-1}), the non-trivial components in ansatz-II may formally be expressed using the ansatz-I under an interchange $\t \leftrightarrow r$ and vice-versa. As a result, the angular momenta are explicitly realized with the ansatz-II at the expense of the conserved charges with ansatz-I. It implies that the conserved charges do interplay among themselves underlying various tunneling vacua on a non-BPS brane. 

%%%%%%%%%%%%%%%%%%%%%%%%%%%%%%%%%%%%%%%%%%%%%%%%%%%%%%%%%%%%%%%%%%%%%%%%%
\subsection{Quantum Kerr geometry: ansatz-II}
%%%%%%%%%%%%%%%%%%%%%%%%%%%%%%%%%%%%%%%%%%%%%%%%%%%%%%%%%%%%%%%%%%%%%%%%%
A vanishing torsion ${\cal H}_3=0$ in the gauge choice ensures a Riemannian curvature in five dimensions on a non-BPS brane in type IIA superstring theory. It implies that the background fluctuations in $B_2$, on a $D_4$-brane, become significant in the gauge choice. In principle, they may have their origin in a higher dimensional $D$-brane with a nontrivial geometric torsion. Nevertheless, the fluctuations do have a source in the local degrees of $B_2$ in the string bulk. In a second gauge choice (\ref{kerr-brane-21}), the non-zero metric components on an effective $D_4$-brane in the prescribed regime, $i.e.\ \rho^4$$>$$>$$(M^2,a^4,b^4)$ with $\rho^2$$>$$(M,a^2,b^2)$, are worked out. The quantum patches on a brane are approximated in the prescribed brane window to yield:
\bea
G_{tt}&=&-\left (1-\frac{2M}{\rho^2}\right ), \nonumber \\
G_{rr}&=&\left (1+ {{{\cal M}^-}\over{r^2\rho^2}}\right )^{-1},\nonumber\\
%G_{\t\t}&=&\rho^2 + {{{\cal M}^+}\over{r^2}} + {{\tri}\over{r^2}}\left (1-G_{rr}^{-1}\right )
%-{{2{\tilde a}^2M}\over{(r^2+a^2)\sin^2\t}}-{{2{\tilde b}^2M}\over{(r^2+b^2)\cos^2\t}}\ ,\nonumber\\
G_{\t\t}&=&\rho^2 + \left [{{{\cal M}^+}\over{r^2}} + {{\tri}\over{r^2}} \left (1-G_{rr}^{-1}\right )
-{{2M}\over{\tri}}\left ( {{{\tilde a}^2(r^2+b^2)}\over{\sin^2\t}} + {{{\tilde b}^2(r^2+a^2)}\over{\cos^2\t}}\right )\right ]\nonumber\\
&=& \rho^2\ +\ {\tilde\rho}_0^2\ ,\nonumber\\
G_{\phi\phi}&=&\left (r^2+a^2 + {{2M}\over{\rho^2}} {{{\tilde a}^2}\over{\sin^2\t}}\right )\sin^2\t\ ,\nonumber\\
G_{\psi\psi}&=&\left (r^2+b^2 + {{2M}\over{\rho^2}} {{{\tilde b}^2}\over{\cos^2\t}}\right )\cos^2\t\ ,\nonumber\\
G_{t\t}&=&{{\sqrt{2M\tri}}\over{r\rho}}\left ({\cal M}\ -\ {{2Mr^2}\over{\tri^2}}\left ( {{{\tilde a}^2(r^2+b^2)}\over{\sin^2\t}} + {{{\tilde b}^2(r^2+a^2)}\over{\cos^2\t}}\right )\right )^{1/2},\nonumber\\
G_{t\phi}&=&-{{2{\tilde a}M}\over{\rho^2}}\ ,\nonumber\\
G_{t\psi}&=&-{{2{\tilde b}M}\over{\rho^2}}\ ,\nonumber\\
G_{\phi\psi}&=&{{2{\tilde a}{\tilde b}M}\over{\rho^2}}\ ,\nonumber\\
G_{\t\phi}&=&{{{\tilde a}\sqrt{2M\tri}}\over{r\rho}}\left ({\cal M}\ -\ {{2Mr^2}\over{\tri^2}}\left ( {{{\tilde a}^2(r^2+b^2)}\over{\sin^2\t}} + {{{\tilde b}^2(r^2+a^2)}\over{\cos^2\t}}\right )\right )^{1/2},\nonumber\\
G_{\t\psi}&=&{{{\tilde b}\sqrt{2M\tri}}\over{r\rho}}\left ({\cal M}\ -\ {{2Mr^2}\over{\tri^2}}\left ( {{{\tilde a}^2(r^2+b^2)}\over{\sin^2\t}} + {{{\tilde b}^2(r^2+a^2)}\over{\cos^2\t}}\right )\right )^{1/2},\label{kerr-brane-22}
\eea
$${\rm where}\quad {\cal M}= \left ({{{\cal M}^+}\over{\tri}}- {{{\cal M}^-}\over{r^2\rho^2}}\right )\ .\qquad\qquad\qquad\qquad\qquad\qquad\qquad\qquad\qquad\qquad {}$$
\begin{figure}
\mbox{
\subfigure{\includegraphics[width=0.45\textwidth,height=0.21\textheight]{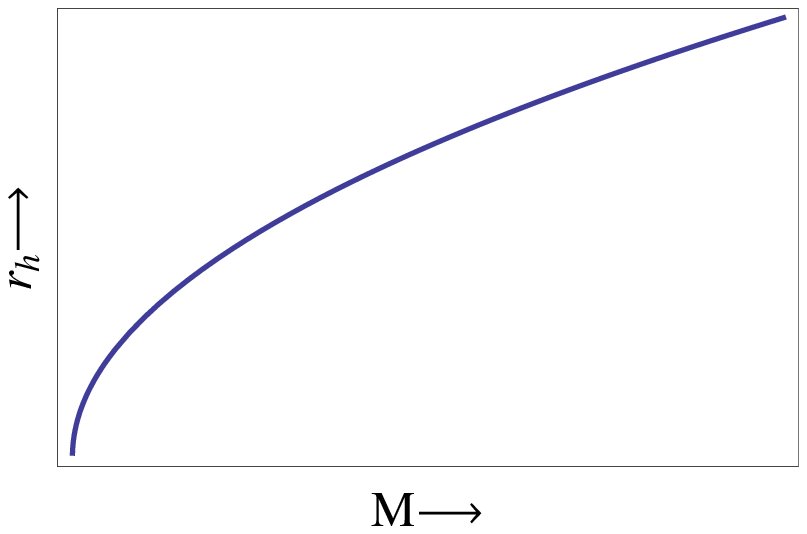}}
\hspace{.15in}
\subfigure{\includegraphics[width=0.45\textwidth,height=0.21\textheight]{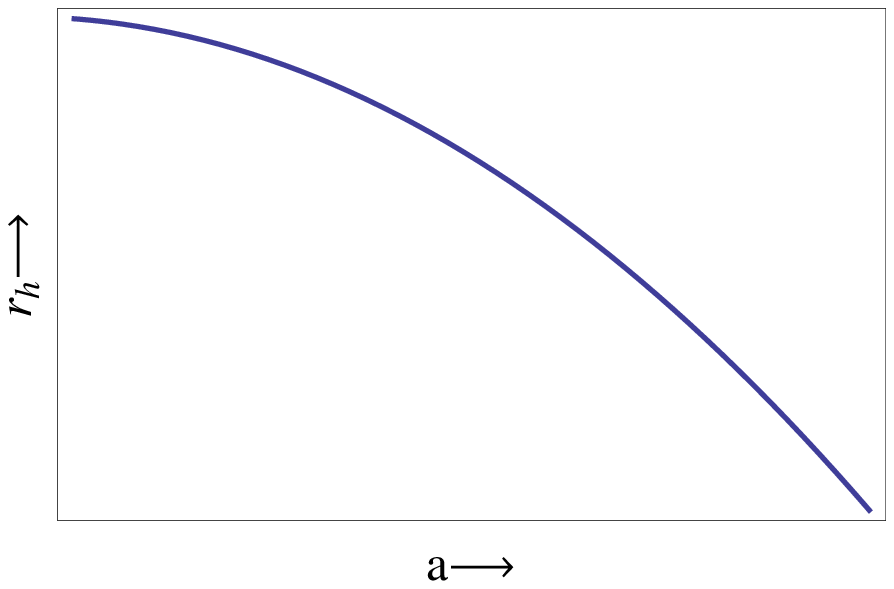}}}
%\mbox{\includegraphics[width=0.53\linewidth,height=0.31\textheight]{Horizon1}}
%\vspace{-.4in}
\caption{\it The event horizon radius $r_h$ is proportional to the mass $M$, for fixed ($a,b$), in a 5D Kerr black hole. However, the horizon radius is inversely proportional to the background parameter: $a$ for fixed ($M,b$) or $b$ for a fixed ($M,a$) and vice-versa. Thus a macroscopic black hole is defined with a large $M$ and for extremely small $a$ and $b$. It is in conformity with the analysis under a variation of an ergo radius in Figure 2 and is consistent with the low energy limit $\tri \rightarrow (r\rho)^2$. Since the causal patches do not reduce in the limit, the variations are characteristic features of a 5D quantum Kerr black hole and its reduced geometry describing a macroscopic black hole.}
\end{figure}

\noindent
The quantum causal patches identify themselves with that in a 5D Kerr black hole. The components of angular velocity are worked out at the event horizon, $r\rightarrow r_+$ and $\t\rightarrow \t_+$. They  are:
\bea
&&\Omega^{\phi}={{-2M\tilde{a}}\over{2{\tilde a}^2M + {\rho_+}^2({r_+}^2+a^2)\sin^2\t_+ }}\ ,\nonumber\\
&&\Omega^{\psi}={{-2M\tilde{b}}\over{2{\tilde b}^2M + {\rho_+}^2({r_+}^2+b^2)\cos^2\t_+ }}\nonumber\\
&&\Omega^{\t}= {{\sqrt{2M}}\over{\rho_+}}\left ({{{\cal M}^+_+ +2Mr_+^2}\over{r_+^2}} - {{2{\tilde a}^2M}\over{(r_+^2+a^2)\sin^2\t_+}} -
{{2{\tilde b}^2M}\over{(r_+^2+b^2)\cos^2\t_+}}\right )^{1/2}\nonumber\\ 
&&\times\ \left (\rho_+^2+ {{{\cal M}^+_+ +2Mr_+^2}\over{r_+^2}} - {{2{\tilde a}^2M}\over{(r_+^2+a^2)\sin^2\t_+}} -
{{2{\tilde b}^2M}\over{(r_+^2+b^2)\cos^2\t_+}}\right )^{-1}.\label{kerr-brane-23}
\eea
The brane geometry (\ref{kerr-brane-22}) possesses a symmetry, independently in its angular coordinates under:
$\left (\t\rightarrow -\t,\ \phi\rightarrow -\phi,\ \psi\rightarrow -\psi\right )$. In addition, the geometry also possesses a symmetry, independently, in some of its parameters under: $\left ( {\sqrt{M}}\rightarrow -{\sqrt{M}},\ a\rightarrow -a,\ b\rightarrow -b\right )$. 
However for $\left ( {\tilde a}\rightarrow -{\tilde a}, {\tilde b}\rightarrow -{\tilde b}\right )$ though $\Omega^\t$ does not change, the remaining two components change: $\Omega^{\phi}\rightarrow -\Omega^{\phi}$ and $\Omega^{\psi}\rightarrow -\Omega^{\psi}$. The quantum geometry on an effective ${\bar D}_4$-brane may be obtained from that on an effective $D_4$-brane 
under $r\rightarrow -r$. In comparision to a Kerr black hole obtained in eq.(\ref{kerr-brane-11}), the barne and anti-brane possess additional patches: $(G_{t\t}, G_{\t\phi}, G_{\t\psi})$. In a global scenario, an absence of RR charge breaks the supersymmetry and leaves behind a non-BPS brane. It is given by
\bea
ds^2&=& -\left(1-\frac{2M}{\rho^2}\right) dt^2\ +\ \left(1-\frac{2M}{\rho^2}+\frac{\left ( a^2\sin^2\t + b^2\cos^2\t\right )r^2 +
a^2b^2}{r^2\rho^2} \right)^{-1} dr^2\ +\ \rho^2 \left ( 1 + {{{\tilde\rho}_0^2}\over{\rho^2}}\right )d\t^2\nonumber\\ 
&&+ \left ( r^2+a^2 + \frac{2{\tilde a}^2M}{\rho^2\sin^2\t }\right) \sin^2\t\ d\phi^2\ +\ \left(r^2+b^2+\frac{2{\tilde b}^2M}{\rho^2\cos^2\t }\right) \cos^2\t \ d\psi^2 \nonumber\\
&&-\ \frac{4{\tilde a}M}{\rho^2}dtd\phi\ -\ \frac{4{\tilde b}M}{\rho^2}dtd\psi\ +\ \frac{4{\tilde a}{\tilde b}M}{\rho^2}d\phi d\psi\ .\label{kerr-brane-24}
\eea
The causal geometric patches precisely correspond to a quantum Kerr black hole (\ref{kerr-brane-8}) obtained with a different ansatz for a two form.
However, the remaining geometric patches appear to differ significantly. The difference modifies the angular momenta of a non extremal quantum Kerr black hole. However, the geometric difference can neither generate an angular momentum nor changes the causal structure in the Kerr vacua. 
One may identify the constants appropriately at the horizon, $i.e.\ {\tilde a}= a\sin^2\t_+$ and ${\tilde b}=b\cos^2\t_+$, in a quantum Kerr black hole (\ref{kerr-brane-24}). With a subtlety, the quantum Kerr becomes 
\bea
ds^2&=& -\left(1-\frac{2M}{\rho^2}\right) dt^2\ +\ \left(1-\frac{2M}{\rho^2}+\frac{\left ( a^2\sin^2\t + b^2\cos^2\t\right )r^2 + a^2b^2}{r^2\rho^2} \right)^{-1} dr^2\nonumber\\
&&+\ \left ( 1 -2M\right )\rho^2\ d\t^2\ +\ 2M\left ( 1 + {1\over{r^2}} + {1\over{\rho^2}} - {{a^2 + b^2}\over{r^2\rho^2}} + {{a^4 \sin^2\t + b^4 \cos^2\t}\over{r^4\rho^2}}\right )\rho^2\ d\t^2\nonumber\\
&&+\ \left ( r^2+a^2 + \frac{2a^2M\sin^2\t}{\rho^2}\right) \sin^2\t\ d\phi^2\ +\ \left(r^2+b^2+\frac{2b^2M \cos^2\t}{\rho^2}\right) \cos^2\t \ d\psi^2\nonumber\\ 
&&-\ \frac{4aM\sin^2\t }{\rho^2}dtd\phi\ -\ \frac{4bM\cos^2\t}{\rho^2}dtd\psi\ +\ \frac{4abM\sin^2\t \cos^2\t}{\rho^2}d\phi d\psi\ .\label{kerr-brane-242}
\eea
The quantum Kerr black holes (\ref{kerr-brane-8}) and (\ref{kerr-brane-24}) differ in their deformation geometries along the $\t$-coordinate. They may be understood as a tunneling vacua in the quantum regime.

%%%%%%%%%%%%%%%%%%%%%%%%%%%%%%%%%%%%%%%%%%%%%%%%%%%%%%%%%%
\subsection{Emergent stringy Kerr black hole in 5D: anstaz-II}
%%%%%%%%%%%%%%%%%%%%%%%%%%%%%%%%%%%%%%%%%%%%%%%%%%%%%%%%%%
The quantum Kerr vacuum (\ref{kerr-brane-242}), in a low energy lmit, reduces to a Kerr black hole \cite{sakaguchi}. The large $r$ limit is defined 
with $r^2>>1$. It may imply $\tri\rightarrow r^2\rho^2$. The metric component $G_{\t\t}$ simplifies drastically in the regime. Interestingly, the 5D Kerr black hole maps to a classical geometry (\ref{kerr-brane-111}) with ${\sqrt{2}}\ \Omega^{\rm II}\rightarrow \Omega^{\rm I}$. The quantum Kerr vacua (\ref{kerr-brane-9}) and (\ref{kerr-brane-24}) tunnels due to the instabilities sourced by the $B_2$-fluctuations 
underlying the fluxes in type IIA/B superstring theory on $S^1$. Nevertheless, a quantum Kerr black hole may be perceived through the tunneling vacua. The energy released during the tunnelings setup a low energy limit leading to Einstein Kerr vacuum. It is given by
\begin{figure}
\mbox{
\subfigure{\includegraphics[width=0.47\textwidth,height=0.22\textheight]{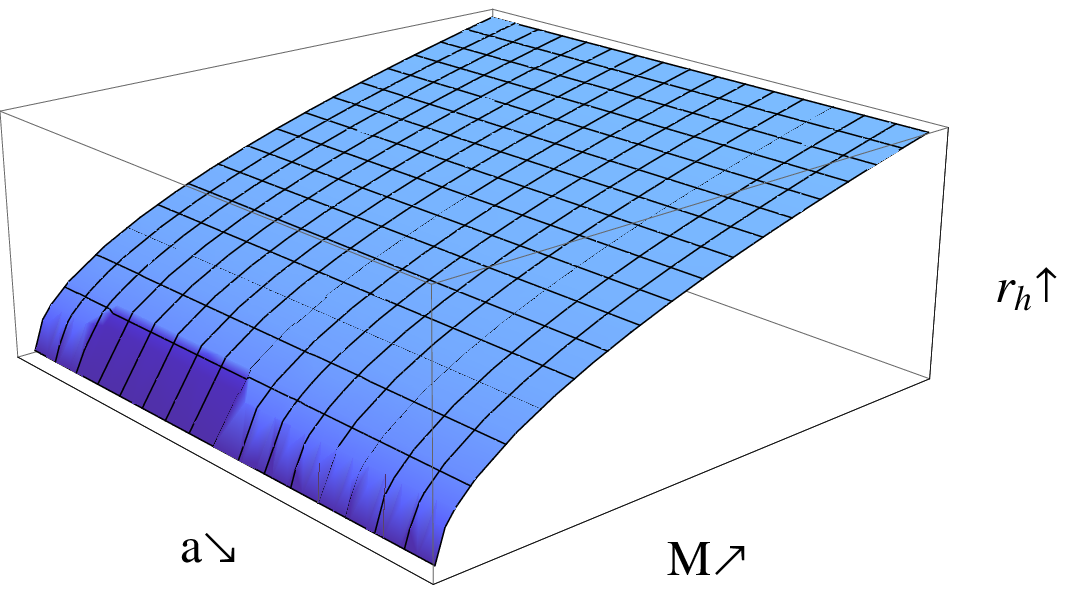}}
\hspace{.08in} 
\subfigure{\includegraphics[width=0.45\textwidth,height=0.22\textheight]{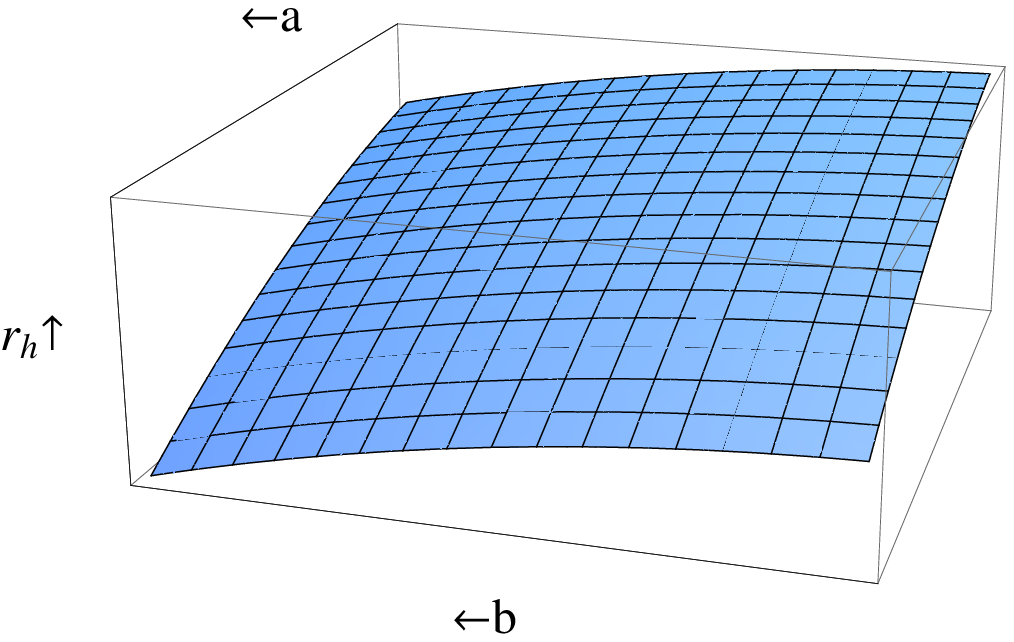}}}
%\mbox{\includegraphics[width=0.53\linewidth,height=0.31\textheight]{Horizon1}}
%\vspace{-.6in}
\caption{\it  Variations of event horizon $r_h$, with (i) ($M,a$) and (ii) $(a,b)$, in a $5D$ Kerr black hole. The 3D plots are in agreement with that for 2D. However, the expansion in the black hole horizon is significantly larger for an increasing $M$ than for the decreasing $a$ or $b$. Thus a macroscopic Kerr black hole, obtained in the low energy limit is defined with a large mass rather than a large $a$ and $b$.}
\end{figure}
\bea
ds^2&=& -\left(1-\frac{2M}{\rho^2}\right) dt^2\ +\ \left(1-\frac{2M}{\rho^2}+\frac{\left ( a^2\sin^2\t + b^2\cos^2\t\right )r^2 + a^2b^2}{r^2\rho^2} \right)^{-1} dr^2\ + \ \rho^2\ d\t^2\nonumber\\
&&+\ \left ( r^2+a^2 + \frac{2a^2M\sin^2\t}{\rho^2}\right) \sin^2\t\ d\phi^2\ +\ \left(r^2+b^2+\frac{2b^2M \cos^2\t}{\rho^2}\right)\cos^2\t \ d\psi^2\nonumber\\ 
&&-\ \frac{4aM\sin^2\t }{\rho^2}\ dtd\phi\ -\ \frac{4bM\cos^2\t }{\rho^2}\ dtd\psi\ +\ \frac{4abM\sin^2\t \cos^2\t}{\rho^2}d\phi d\psi\ .\label{kerr-brane-243}
\eea
On the other hand, the quantum Kerr black hole (\ref{kerr-brane-24}) when viewed through a $\t$-sliced (say $\t=\pi/4$) geometry 
for ${\tilde a}=(a/2)$ and ${\tilde b}=(b/2)$ reduces precisely to a $\t$-sliced typical Kerr black hole in five dimensions. It is given by
\bea
ds^2&=& -\left (1- {{2M}\over{\rho^2}}\right )dt^2 + \left ( 1-{{2M}\over{\rho^2}} + {{(a^2+b^2)r^2+2a^2b^2}\over{2r^2\rho^2}}\right )^{-1}dr^2\ 
-\ {{2aM}\over{\rho^2}}\ dt d\phi\ -\ {{2bM}\over{\rho^2}}\ dt d\psi\nonumber\\
&&+\ {1\over2}\left (r^2+a^2+ {{a^2M}\over{\rho^2}}\right )d\phi^2\ +\ {1\over2}\left ( r^2+b^2+ {{b^2M}\over{\rho^2}}\right )d\psi^2 
\ .\label{kerr-brane-25}
\eea
Thus, a $\t$-sliced five dimensional Kerr black hole is common to both on a non-BPS brane and in Einstein gravity. It may imply that a quantum Kerr black hole (\ref{kerr-brane-24}) does not access the full quantum vacuum as expected from a perturbative quantum theory of gravity.  

%%%%%%%%%%%%%%%%%%%%%%%%%%%%
\section{Concluding remarks}
%%%%%%%%%%%%%%%%%%%%%%%%%%%%
A generalized curvature theory, primarily sourced by a two form in a $U(1)$ gauge theory, on a $D_4$-brane is revisited to obtain some of the quantum Kerr vacua in five dimensions. Interestingly the non-extremal black holes, underlying the Kerr geometries, are emergent on a pair of BPS brane and anti-BPS brane in a global scenario. In other words, the Kerr geometries were constructed in a non-perturbative framework, underlying a geometric torsion, on a non BPS brane. Since a torsion is intrinsic to the framework, the angular velocity sourced by a torsion on a brane is nullified by that on an anti-brane. As a result, a non BPS brane geometries have been shown to describe some of the established Einstein vacua underlying a superstring theory.

\sp
\noindent
In particular, the five dimensional emergent Kerr geometries purely sourced by the background fluctuations in $B_2$ on a non-BPS brane were constructed. The gauge choice freezes the local degrees in torsion, which is described by a five dimensional generalized curvature theory on a pair of $(D{\bar D})_4$-brane. A vanishing torsion may alternately be viewed as a vanishing energy momentum tensor in a non-linear $U(1)$ gauge theory. Thus, a two form fluctuations, on a non BPS brane enhance the possibility to map the emergent gravity in a non-perturbative framework to the classical vacuum in Einstein gravity. Nevertheless, the background fluctuations in five dimensions may be understood in presence of an electromagnetic field $F_2$, which incorporates local degrees into a non-linear global description. The local $F_2$ may be gauged away in the framework. In addition, the background fluctuations may have their origin in a dynamical two form, and may describe a propagating torison, in higher dimensions. A two form gauge transformations do not make the emergent geometries unique. In fact, there are a very large number of emergent vacua. They have been qualitatively argued to describe the landscape vacua in string theory. Interestingly, an emergent geometry possesses two angular momenta. The quantum Kerr vacua in a low energy limit have been shown to describe a typical Kerr black hole in Einstein vacuum. Interestingly, a quantum Kerr-Newman black hole in 4D on 
an effective $D_4$-brane on $S^1$ is in progress by the authors.

\sp
\noindent
Our investigations with a geometric torsion in a non-perturbative curvature theory may be viewed as a construction under the conjectured M-theory in eleven dimensions. Further analysis may suggest that a two form in a non-linear $U(1)$ gauge theory presumably provide a clue to the source of dark energy. It is plausible to believe that the dark energy in a gravity theory may have its origin in a geometric torsion sourced by a two form. 
In the context, the wall crossing formula relating a single centered black hole to a multi-centred may be applied to the tunneling vacua on a
non-BPS brane to compute the black hole entropy exactly. Presumably, they would like to enhance our understanding on the strong-weak coupling duality leading to non-perturbative world.
%%%%%%%%%%%%%%%%%%%%%%%%%%
\section*{Acknowledgments}
%%%%%%%%%%%%%%%%%%%%%%%%%%
S.S. acknowledges UGC and A.K.S. acknowledges CSIR for their fellowship. The work of S.K. is partly supported by a research grant-in-aid under the Department of Science and Technology, Government of India.

%**********************************************************%
\def\anp{Ann. of Phys.}
\def\cmp{Comm.Math.Phys}
\def\prl{Phys.Rev.Lett.}
\def\jmp{J.Math.Phys.}
\def\prd#1{{Phys.Rev.}{\bf D#1}}
\def\jhep{JHEP\ {}}{}
\def\cqg#1{{Class.\& Quant.Grav.}}
\def\plb#1{{Phys. Lett.} {\bf B#1}}
\def\npb#1{{Nucl. Phys.} {\bf B#1}}
\def\mpl#1{{Mod. Phys. Lett} {\bf A#1}}
\def\ijmpa#1{{Int.J.Mod.Phys.}{\bf A#1}}
\def\mpla#1{{Mod.Phys.Lett.}{\bf A#1}}
\def\rmp#1{{Rev. Mod. Phys.} {\bf 68#1}}

%**********************************************************%

\end{document}